\documentclass [12pt,a4paper]{article}
\usepackage{amssymb, theorem}
\newtheorem{Thm}{Theorem}
\newtheorem{lemma}{Lemma}
\theorembodyfont{\rm}
\newtheorem{pl}{Example}
\newtheorem{Q}{Question}
\def\proof{{\it Proof: }}

\def\qed{\nobreak\hfill $\square$}
\def\<{\langle}
\def\>{\rangle}

\def\iH{{\cal H}}

\def\iA{{\cal A}}
\def\iB{{\cal B}}
\def\iC{{\cal C}}

\def\iU{{\cal U}}

\def\iM{{\cal M}}
\def\iN{{\cal N}}
\def\iE{{\cal E}}
\def\iF{{\cal F}}
\def\perpp{\perp_0}

\def\Ms{M_s(\bbbc)}
\def\iso{\simeq}

\def\im{{\rm i}}
\def\ot{\otimes}

\def\bbbc{{\mathbb C}}

\def\Tr{\mbox{Tr}\,}
\def\pont{{\, \cdot \,}}

\begin{document}
\ \vskip 1cm 
\centerline{\LARGE {\bf Complementarity in Quantum Systems}}
\bigskip\bigskip
\bigskip
\centerline{\large D\'enes Petz \footnote{E-mail: petz@math.bme.hu.
Partially supported by the Hungarian Research Grant OTKA T032662.} }
\bigskip
\centerline{Alfr\'ed R\'enyi Institute of Mathematics}
\centerline{Hungarian Academy of Sciences}
\centerline{ H-1053 Budapest, Re\'altanoda u. 13-15, Hungary }
\bigskip
\bigskip\bigskip

\begin{quote}
{\bf Abstract:} 
Reduction of a state of a quantum system to a subsystem gives partial 
quantum information about the true state of the total system.  Two 
subalgebras $\iA_1$ and $\iA_2$ of $B(\iH)$ are called complementary 
if the traceless subspaces of $\iA_1$ and $\iA_2$ are orthogonal (with 
respect to the Hilbert-Schmidt inner product). When both subalgebras 
are maximal Abelian, then the concept reduces to complementary observables 
or mutually unbiased bases. In the paper several characterizations of 
complementary subalgebras are given in the general case and several 
examples are presented. For a 4-level quantum system, the structure 
of complementary subalgebras can be described very well, the Cartan 
decomposition of unitaries plays a role. It turns out that a measurement 
corresponding to the Bell basis is complementary to any local measurement 
of the two-qubit-system.

\end{quote}

The study of complementary observables goes back to early quantum mechanics.
Position and momentum are the typical examples of complementary observables
and the main subject was the joint measurement and the uncertainty \cite{BL,
BS}. In the 
setting of finite dimensional Hilbert space and in a mathematically rigorous 
approach, the paper \cite{Sch} of Schwinger might have been the first in 1960. 
The goal of that paper is the finite dimensional approximation of the 
canonical commutation relation. An observable of a finite system can be 
identified with a basis of the Hilbert space through the spectral theorem 
\cite{Ac} and instead of complementarity the expression ``mutually unbiased''
became popular \cite{WF}. The maximum number of mutually unbiased bases is
still and open question \cite{MUB}, nevertheless such bases are used in 
several contexts, state determination, the ``Mean King's problem'', 
quantum cryptography etc. \cite{Iv, Kimura, Bruss}. 

Motivated by the frequent use of mutually unbiased bases and complementary 
reductions of two qubits \cite{PHSz, nofive}, the goal of this paper is a
general study of complementary subalgebras. The particular case, when the
subalgebras are maximal Abelian, corresponds to complementary observables,
or mutually unbiased bases. This case has been studied in the literature
by many people. If the reduction of a quantum state to a subalgebra is
known to us, then this means a partial information about the state. The
concept of complementarity of two subsystems means heuristically that
the partial information provided jointly by the two subsystems is the 
largest when it is compared with the information content of the two
subsystems \cite{WF}.

The paper is organized in the following way. First the entropic uncertainty
relation of Maasen and Uffink is reviewed as a motivation for the concept
of complementarity (of observables or basis). Then the complementarity of
observables is reformulated in terms of commutative subalgebras. This
reformulation leads to the complementarity of more general subalgebras
(corresponding to a subsystem of a quantum system). It turns out that
complementarity is a common generalization of the ordinary tensor product 
and the twisted fermionic tensor product. When two subalgebras are unitarily
equivalent, complementarity can be read out from the unitary when it is
viewed as a block-matrix. A modification of the construction of complementary 
bases (going back to Schwinger) yields examples of complementary subalgebras
in arbitrary dimension. The maximal number of complementary subalgebras remains an
open question, however, the case of 4-level quantum system is analyzed in details. 
It turns out that a measurement corresponding to the Bell basis is complementary 
to any local measurement of the two-qubit-system.
  
\section{Complementary observables}

Let $A$ and $B$ be two self-adjoint operators on a finite dimensional Hilbert
space. If $A=\sum _i\lambda_i^AP_i^A$ and $B=\sum_i\lambda_i^BP_i^B$ are their
spectral decompositions, then
\[
H(A,\varphi)=\sum_i\eta (\varphi(P_i^A))
\quad {\rm and} \quad
H(B,\varphi)=\sum_i\eta (\varphi(P_i^B))
\]
are the {\bf entropies} of $A$ and $B$ in a state $\varphi$. ($\eta (t)$
is the function $-t \log t$.)

Assume that the eigenvalues of $A$ and $B$ are free from multiplicities. If 
these observables share a common eigenvector and the system is prepared in 
the corresponding state, then the measurement of both $A$ and $B$ leads to 
a sharp distribution and one cannot speak of uncertainty. In order to exclude 
this case, let $(e _i)$ be an orthonormal basis consisting of eigenvectors 
of $A$, let $(f_i)$ be a similar basis for $B$ and we suppose that
\begin{equation}
c^2:=\sup \,\{\vert \< e_i,f_j\rangle\vert^2\colon i,j\}
\label{(16.9)}
\end{equation}
is strictly smaller than 1. Then $H(A,\varphi)+H(B,\varphi)>0$ for
every pure state $\varphi$. Since the left-hand-side is concave in 
$\varphi$, it follows that $H(A,\varphi)+H(B,\varphi)>0$ for any state
$\varphi$. This inequality is a sort of uncertainty relation. The lower
bound was conjectured in \cite{Kr} and proven by Maasen and Uffink in 
\cite{M-U}.

\begin{Thm}\label{thm:16.2}
With the notation above the uncertainty relation
\[
H(A,\varphi)+H(B,\varphi)\ge - 2 \log c
\]
holds.
\end{Thm}

Let $n$ be the dimension of the underlying Hilbert space. We may assume that
$\varphi$ is a pure state corresponding to a vector $\Phi$. Then 
$\varphi(P_i^A)=|\<e_i,\Phi\>|^2$ and $\varphi(P_i^B)=|\<f_i,\Phi\>|^2$. 

The $n\times n$ matrix $T_{i,j}:=(\< e_i,f_j\rangle )_{i,j}$ is
unitary and $T$ sends the vector
\[
f:= (\< e_1,\Phi\rangle ,\< e_2,\Phi\rangle
,\dots,\< e_n,\Phi\rangle)
\]
into
\[
Tf=(\< f_1,\Phi\rangle ,\< f_2,\Phi\rangle
,\dots,\< f_n,\Phi\rangle) \, .
\]
The vectors $f$ and $Tf$ are elements of $\bbbc ^n$ and this space may be
endowed with different $L^p$ norms. Using interpolation theory we shall
estimate the norm of the linear transformation $T$ with respect to
different $L^p$ norms. Since $T$ is a unitary
\[
\Vert g\Vert_2=\Vert Tg\Vert_2\qquad (g\in\bbbc ^n)\,.
\]
With the notation (\ref{(16.9)}) we have also
\[
\Vert Tg\Vert_\infty\le c\Vert g\Vert_1\qquad (g\in\bbbc ^n)\,.
\]
Let us set
\[
N(p,p')=\sup \{\Vert Tg\Vert_p/\Vert g\Vert_{p'}\colon g\in\bbbc ^n,\quad
g\ne 0\}
\]
for $1\le p\le\infty$ and $1\le p'\le\infty$. The {\bf Riesz--Thorin convexity
theorem} says that the function
\begin{equation}
(t,s)\mapsto\log N(t^{-1},s^{-1})\label{(16.10)}
\end{equation}
is convex on $\lbrack 0,1\rbrack \times \lbrack 0,1\rbrack$ (where $0^{-1}$ is
understood to be $\infty$). Application of convexity of (\ref{(16.10)}) on the
segment $\lbrack (0,1),(1/2,1/2)\rbrack$ yields
\[
\Vert Tg\Vert_{2/\lambda}\le c^{1-\lambda}\Vert g\Vert_\mu\qquad
 (g\in\bbbc ^n)\, ,
\]
where $0<\lambda <1$ and $\mu=(1-\lambda/2)^{-1}$. This is rewritten by means
of a more convenient parameterization in the form
\[
\Vert Tg\Vert_p\le c^{1-2/p}\Vert g\Vert_q\qquad (g\in\bbbc ^n)\, ,
\]
where $2\le p<\infty$ and $p^{-1}+q^{-1}=1$. Consequently
\begin{equation}
\log\Vert Tf\Vert_p\le \Big(1-{2\over p}\Big)\log c+\log\Vert
f\Vert_q\,.\label{(16.11)}
\end{equation}
One checks easily that
\[
 {d\log\Vert Tf\Vert_p\over dp}\Big\vert_{p=2}=- {1\over 4}H(B,\varphi)
\quad
{\rm and} \quad
 {d\log\Vert f\Vert_q\over dp}\Big\vert_{p=2}= {1\over 4}H(A,\varphi)\,.
\]
Hence dividing (\ref{(16.11)}) by $p-2$ and letting $p\searrow 2$ we obtain
\[
\textstyle{- {1\over 4}H(B,\varphi)\le {1\over 2}\log c+ {1\over
4}H(A,\varphi) }
\]
which proves the theorem for a pure state.

Concavity of the left hand side of the stated inequality in $\varphi$ ensures
the lower estimate for mixed states. \qed

The theorem can be formulated in an algebraic language. Let $\iA$ and $\iB$
be maximal Abelian subalgebras of the algebra $M_n(\bbbc)$ of $n \times n$
matrices. Set
\begin{equation}
c^2:=\sup \,\{\Tr PQ: P\in \iA, Q\in \iB \mbox{\ are\ minimal\ projections} \}.
\label{(16.92)}
\end{equation}
The theorem tells that
\begin{equation}\label{E:MU}
H(\varphi|\iA)+H(\varphi|\iB) \ge - 2 \log c\,.
\end{equation}
Both the definition of $c$ and the statement are formulated without the
underlying Hilbert space. 

\begin{Q} 
Can we make the proof of (\ref{E:MU}) without using the Hilbert space?
\end{Q}

Let $A$ and $B$ self-adjoint operators with eigenvectors $(e_j)$ and
$(f_i)$, respectively and let $\varphi$ be the pure state corresponding to
$e_1$. Then $H(A,\varphi)=0$ and $H(B,\varphi)=\log n$. Hence this example
shows that the lower bound for the entropy sum in Theorem~\ref{thm:16.2} is
sharp. If  (\ref{(16.12)}) holds then the pair $(A,B)$ of observables are 
called {\bf complementary} \cite{Ac}. According to another terminology, the 
bases $(e_j)_j$ and $(f_k)_k$ are called {\bf mutually unbiased} if 
(\ref{(16.12)}) holds. Mutually unbiased bases appeared in a different 
setting in the paper \cite{Iv, WF}, where state determination was discussed.

The lower bound in the uncertainty (\ref{E:MU}) is the largest if $c^2$ is the 
smallest. Since $n^2 c^2 \ge n$, the smallest value of $c^2$ is $1/n$. This happens
if and only if
\begin{equation}
\vert\< e_j,f_k\rangle\vert^2=n^{-1}\qquad
(j,k=1,2,\dots ,n)\,,\label{(16.12)}
\end{equation}
that is, the two bases are mutually unbiased. This is an extremal property of
the mutually unbiased bases. The largest lower bound is attained if $\phi$
is a vector state generated by one of the basis vectors.

The complementarity of observables is also the property of the spectral 
measures associated with them. Therefore the extension to POVM's is natural. 
For a POVM $\iE\equiv (E_i)_i$ and for a unit vector $\Phi$, we define an 
entropy quantity as
$$
H(\iE, \Phi)=\sum_i \Tr \eta(\<\Phi, E_i\Phi\>).
$$

Let $\iE=(E_i)_i$ and $\iF=(F_j)_j$ be POVM's on a Hilbert space $\iH$ 
and $\Phi \in \iH$ be a unit vector. Then the inequality
$$
H(\iE,\Phi)+H(\iF, \Phi) \geq -2 \log \sup\left\{\frac{|\<\Phi, E_iF_j\Phi\>|}
{\<\Phi, E_i\Phi\>\<\Phi, F_j\Phi\>} : i,j\right\}
$$
holds and was proven in \cite{Kr-Pa}. This estimate is essentially different 
from the uncertainty relation of Theorem \ref{thm:16.2}. The lower bound here 
depends on the vector $\Phi$.

\begin{Q}  
What is the lower bound if $F_{j}\Phi=\Phi$ for a certain $j$? 
\end{Q}

The uncertainty relation in Theorem \ref{thm:16.2} is for two observables. 
Assume that 
$n+1$ pairwise unbiased observables $A_1,A_2,\dots, A_{n+1}$ are measured
when the system is in state $\varphi$. Sanchez \cite{Sanchez} proved that
\begin{equation} \label{eq:san}
\sum_{k=1}^{n+1} H(A_k,\varphi) \geq (n+1) \log \frac{1}{2}(n+1).
\end{equation}

\section{Complementary subalgebras}

There is an obvious correspondence between bases and maximal Abelian 
subalgebras. Given a basis, the linear operators diagonal in this basis
form a maximal Abelian subalgebra, conversely if $|e_i\>\<e_i|$ are 
minimal projections in a maximal Abelian subalgebra, then $(|e_i\>)_i$ is
a basis. Parthasarathy characterized mutually unbiased bases through
the corresponding maximal Abelian subalgebras. 
 
\begin{Thm}\label{T:Par}
Let $\iA_1$ and $\iA_2$ be maximal Abelian subalgebras of $M_n(\bbbc)$.
Then the following conditions are equivalent:
\begin{enumerate}
\item[(i)] 
If $P \in \iA_1$ and $Q \in \iA_2$ are minimal projections,
then $\Tr PQ=1/n$.
\item [(ii)]
The subspaces $\iA_1 \ominus \bbbc I$ and  $\iA_2 \ominus \bbbc I$
are orthogonal in $M_n(\bbbc)$.
\end{enumerate}
\end{Thm}

Mutually unbiased bases are interesting from many point of view \cite{Kr, 
BSTW} and the maximal number of such bases is not completely
known \cite{Zych}. 

Subalgebras cannot be orthogonal. We say that the subalgebras $\iA_1$ and 
$\iA_2$ are {\bf quasi-orthogonal} if  $\iA_1 \ominus \bbbc I$ and  
$\iA_2 \ominus \bbbc I$ are orthogonal. If $\iA_1$ and  $\iA_2$ are 
quasi-orthogonal, then we use the notation $\iA_1 \perpp \iA_2$. This 
terminology is mathematically very natural. However, from the view point
of quantum mechanics, {\bf complementarity} could be a better expression. If
the subalgebras are maximal Abelian, then they correspond to observables and
quasi-orthogonality of the subalgebras is equivalent to complementarity of the
observables \cite{Ac, Kr, OP}. 

We consider subalgebras of $\iA \equiv M_n(\bbbc)$ such that their minimal 
projections have the same trace. Such subalgebras will be called {\bf homogeneous}.
A maximal Abelian subalgebra and a subalgebra isomorphic to a full matrix algebra 
are homogeneous. (Recall that if  $M_r(\bbbc) \iso \iA_0  \subset M_n(\bbbc)$, then 
up to isomorphism $M_n(\bbbc)=\iA_0 \ot \Ms$ and $rs=n$.) 

The Hilbert-Schmidt inner product will be used in the form
\begin{equation}\label{E:HS}
\< A,B\>=\tau (A^*B)\, ,
\end{equation}
where $\tau$ is the normalized trace. (Of course, orthogonality does not depend on
the normalization of the trace.) 

The next statement is an extension of Parthasarathy's  result.

\begin{Thm}\label{T:uj}
Let $\iA_1$ and $\iA_2$ be homogeneous subalgebras of $M_n(\bbbc)$. Then the 
following conditions are equivalent:
\begin{enumerate}
\item[(i)] 
If $P \in \iA_1$ and $Q \in \iA_2$ are minimal projections,
then $\tau (PQ)=\tau(P)\tau)Q)$.
\item [(ii)]
The subalgebras $\iA_1$ and  $\iA_2$ are quasi-orthogonal in $M_n(\bbbc)$.
\item [(iii)]  $\tau(A_1 A_2)=\tau(A_1)\tau(A_2)$ if $A_1 \in \iA_1$,  $A_2 \in 
\iA_2$.
\item [(iv)]  If $E_1:\iA \to \iA_1$ is the trace preserving conditional expectation, then
$E_1$ restricted to $\iA_2$ is a linear functional (times $I$). 
\end{enumerate}
\end{Thm}

\proof
Note that $\tau((A_1 - I \tau (A_1))(A_2 - I \tau (A_2)))=0$ and $\tau(A_1 A_2))=
\tau(A_1) \tau(A_2)$ are equivalent. If they hold for minimal projections, they
hold for arbitrary operators as well. Moreover, (iv) is equivalent to the property
$\tau(A_1 E_1(A_2))=\tau(A_1 (\tau(A_2))I)$ for every $A_1 \in \iA_1$ and  
$A_2 \in \iA_2$. \qed

Condition (iii) is the independence of the subalgebras with respect to the
tracial state. (Property (iii) is much weaker than the statistical
independence of subalgebras, cf. \cite{Summers}.) 
Condition (iv) can be formulated as the following commuting square:
\begin{figure}[!ht]
\unitlength=1mm
\begin{center}
\begin{picture}(60,30)
\put(30,1){\makebox(0,0)[cc]{$\bbbc I$}}
\put(30,29){\makebox(0,0)[cc]{$M_n(\bbbc)$}}
\put(2,15){\makebox(0,0)[cc]{$\iA_{1}$}}
\put(58,15){\makebox(0,0)[cc]{$\iA_{2}$}}
\put(25,4){\vector(-2,1){17}}
\put(24,26){\vector(-2,-1){16}}
\put(52,12.5){\vector(-2,-1){17}}
\put(52,18){\vector(-2,1){16}}
\put(45,6.5){\makebox(0,0)[lc]{$\tau$}}
\put(44,25.5){\makebox(0,0)[lc]{$\,$}}
\put(14,6.5){\makebox(0,0)[rc]{$\,$}}
\put(16,25.5){\makebox(0,0)[rc]{$E_1$}}
\end{picture}
\end{center}
\caption{Commuting square, $E_{1}|\iA_{2}=\tau(\pont)I$.}
\end{figure}

Assume that $\iA_1,\iA_2,\dots, \iA_m$ are complementary maximal 
Abelian subalgebras of $M_n(\bbbc)$. Since the dimension of $\iA_a \ominus 
\bbbc I$ is $n-1$, the inequality $n^2 -1 \ge m(n-1)$ holds. This implies
that $m \le n+1$.

\section{Mutually unbiased bases}

Let $\iH$ be an $n$-dimensional Hilbert space. The standard construction 
of mutually unbiased bases goes through unitary operators. Assume that
$U_0\equiv I, U_1,\dots,U_{n^2-1}$ is a family of unitaries such that
$$
\Tr U_j^* U_k=0 \quad \mbox{for} j\ne k.
$$
(In other words, $n^{-1/2}U_i$ is an orthonormal basis in $B(\iH)$,
$0 \le i \le n^2-1$.) 

\begin{pl}\label{pl:Sch}
Let  $e_0, e_1,\dots, e_{n-1}$ be a basis and let $X$ be the unitary 
operator permuting the basis vectors cyclically:
$$
Xe_i=\left\{ \begin{array}{ll}
e_{i+1}\quad & \hbox{if } 0 \le i \le n-2,
\\
e_0 &\hbox{if } i=n-1.\end{array} \right.
$$
Let $q:=e^{\im 2\pi/n}$ and define another unitary by $Ye_i=q^i e_i$.  
It is easy to check that $YX=qXY$ or more generally the commutation
relation
\begin{equation}
Z^kX^\ell= q^{k\ell}X^\ell Z^k
\end{equation}
is satisfied. For $S_{j,k}\:=Z^j X^k$, we have
$$
S_{j,k}=\sum_{m=0}^{n-1} q^{mj} |e_m\>\<e_{m+k}|\quad \mbox{and}\quad
S_{j,k}S_{u,v}=q^{ku}S_{j+u,k+v},
$$
where the additions $m+k, j+u, k+v$ are understood modulo $n$. 
(What we have is a finite analogue of the Weyl commutation relation, 
see \cite{Sch}.) Since $\Tr S_{j,k}=0$ when at least one of $j$ and $k$
is not zero, the unitaries
$$
\{S_{j,k}\, :\, 0 \le j,k \le n-1\}
$$
are pairwise orthogonal.

Note that $S_{j,k}$ and $S_{u,v}$ commute if $ku=jv$ mod $n$.  

In the case of $n=2$, $X=\sigma_1$ and $Z=\sigma_3$. (This fact motivated
our notation.) \qed
\end{pl}

Assume that $\iU_1, \iU_2,\dots, \iU_{n+1}$ is a partition of the
set $\{U_1,U_2,\dots,U_{n^2-1}\}$ such that $\# (\iU_j)=n-1$ and $\iU_j$
consists of commuting unitaries. Then the maximal Abelian subalgebras
$\iA_i$ generated by $\iU_i$ are pairwise complementary. (Note that $\iA_i$ 
is the linear span of $I$ and $\iU_i$.) The remaining question is about
the construction of the partition satisfying the requirements.

\begin{pl}
Consider a 4-level quantum system and view $B(\iH)$ as $M_2(\bbbc)\ot 
M_2(\bbbc)$. The unitaries $\sigma_i\ot \sigma_j$ form a basis and the
partition
\begin{eqnarray*}
&&\sigma_0 \ot \sigma_0, 
\cr &&
\sigma_0 \ot \sigma_1, \sigma_1 \ot \sigma_0, \sigma_1 \ot \sigma_1,
\cr &&
\sigma_0 \ot \sigma_2, \sigma_2 \ot \sigma_0, \sigma_2 \ot \sigma_2,
\cr &&
\sigma_0 \ot \sigma_3, \sigma_3 \ot \sigma_0, \sigma_3 \ot \sigma_3,
\cr &&
\sigma_1 \ot \sigma_2, \sigma_2 \ot \sigma_3, \sigma_3 \ot \sigma_1,
\cr &&
\sigma_1 \ot \sigma_3, \sigma_2 \ot \sigma_1, \sigma_3 \ot \sigma_2,
\end{eqnarray*}
determines 5 mutually unbiased bases.
\end{pl}

The terminology of ``{\it mutually unbiased bases}'' was introduced 
in \cite{WF}, where it was showed that the corresponding measurements  
``{\it provide an optimal means of determining an ensemble's state}''.
A slightly different extremal property of mutually unbiased bases is
discussed in \cite{PHM}.

\section{More about complementary subalgebras}

If the pairwise complementary subalgebras $\iA_1,\iA_2,\dots, \iA_{r}$ are 
given and they span the whole algebra $\iA$, then any operator is the sum of 
the components in the subspaces $\iA_a \ominus  \bbbc I$ ($1 \le a \le r$) 
and $\bbbc I$:
\begin{equation}
A= - \tau(A)(r-1)I +\sum_{i=1}^r E_i(A)\, ,
\end{equation}
where $E_i:\iA \to \iA_i$ is the trace preserving conditional expectation
(which is nothing else but the orthogonal projection with respect to the 
Hilbert-Schmidt inner product).

\begin{pl}
Let $\iA_1$ be the subalgebra $\bbbc I \otimes M_r(\bbbc)$  and
$\iA_2$ be the subalgebra $M_p(\bbbc)\otimes \bbbc I$ of $M_p(\bbbc)\otimes 
M_r(\bbbc)$. Then $\iA_1$ and  $\iA_2$ are complementary.

For $p=r=2$ we have
$$
\iA_1=
\left\{\left[\begin{array}{rrrr}
a & b & 0 & 0\\
c & d & 0 & 0\\
0 & 0 & a & b\\
0 & 0 & c & d\\
\end{array}\right]:a,b,c,d \in \bbbc\right\}, \quad
\iA_2=
\left\{\left[\begin{array}{rrrr}
a & 0 & b & 0\\
0 & a & 0 & b\\
c & 0 & d & 0\\
0 & c & 0 & d\\
\end{array}\right]:a,b,c,d \in \bbbc \right\}\,.
$$
For the unitary
\begin{equation}\label{E:LR}
U:=\left[\begin{array}{rrrr}
1 & 0 & 0 & 0\\
0 & 0 & 1 & 0\\
0 & 1 & 0 & 0\\
0 & 0 & 0 & 1\\
\end{array}\right],
\end{equation}
we have $U(I \ot A)U^*=A \ot I$ for every $A \in M_2(\bbbc)$. \qed
\end{pl}

\begin{pl}\label{Pl:2}
Try to find a unitary
$$
W:=\left[\begin{array}{rr}
W_1 & W_2 \\ W_3 & W_4 \end{array}\right]
$$
such that the subalgebra
$$
W\left[\begin{array}{rr} A & 0 \\ 0 & A \end{array}\right]W^*
\qquad (A \in  M_2(\bbbc))
$$
is complementary to $I \ot M_2(\bbbc)$. Assume that $\Tr B=0$. Then the
orthogonality
$$
W\left[\begin{array}{rr} A & 0 \\ 0 & A \end{array}\right]W^* \perp
\left[\begin{array}{rr} B & 0 \\ 0 & B \end{array}\right]
$$
means that 
$$
\Tr (W_1AW_1^*+W_2AW_2^*+W_3AW_3^*+W_4AW_4)B=0.
$$
This holds for every $B$ if and only if 
$$
W_1AW_1^*+W_2AW_2^*+W_3AW_3^*+W_4AW_4
$$
is a multiple of the identity. Therefore, the sufficient and necessary
condition is the following:
\begin{equation}\label{E:17}
W_1AW_1^*+W_2AW_2^*+W_3AW_3^*+W_4AW_4^*=(\Tr A)I \qquad (A \in  M_2(\bbbc).
\end{equation}

For the unitary
\begin{equation}\label{E:Paulik}
W:=\frac{1}{\sqrt{2}}\left[\begin{array}{rr}
I & \sigma_3 \\ \sigma_1 &  \im \sigma_2 \end{array}\right]
\end{equation}
the condition holds. ($\sigma_i$'s are the Pauli matrices.) One computes
that
$$
W(I\ot \sigma_1)W^*= \sigma_1 \ot I, \quad W(I\ot \sigma_2)W^*= \sigma_2 \ot 
\sigma_3, \quad  W(I\ot \sigma_3)W^*= \sigma_3 \ot \sigma_3.
$$

We obtained an algebra determined by a Pauli triplet consisting elementary
tensors of Pauli matrices.  \qed
\end{pl}

The previous example can be generalized.

\begin{Thm}\label{T:2}
Let $W= \sum_{ij=1}^n E_{ij} \ot W_{ij} \in M_n(\bbbc) \ot M_m(\bbbc)$ be 
a unitary, where $E_{ij}$ are the matrix units in $ M_n(\bbbc)$ and
$W_{ij} \in M_m(\bbbc)$. The subalgebra $W(\bbbc  I\ot M_m(\bbbc))W^*$ 
is complementary to $\bbbc  I\ot M_m(\bbbc)$ if and only if 
$$
\frac{m}{n}\sum_{i,j=1}^n |W_{ij}\> \<W_{ij}|=I.
$$
When $n=m$ this condition means that  $\{W_{ij}:1 \le i,j \le n\}$ 
is an orthonormal basis in $ M_n(\bbbc)$ (with respect to the inner product
$\< A,B\>=\Tr A^*B$).
\end {Thm}

\proof
Assume that $A, B \in M_m(\bbbc)$ and  $\Tr B=0$. Then the condition
$$
W(I \ot A^*)W^* \perp (I \ot B)
$$
is equivalently written as
$$
\Tr W(I \ot A)W^* (I \ot B)= \sum_{i,j=1}^n \Tr W_{ij} A W_{ij}^* B=0.
$$
Putting $B-(\Tr B)I_m /m$ in place of $B$, we get
$$
\sum_{i,j=1}^n \Tr W_{ij} A W_{ij}^* B=\frac{1}{m}\Tr B 
\sum_{i,j=1}^n \Tr W_{ij} A W_{ij}^*\, .
$$ 
for every $B \in M_m(\bbbc)$. Since $W$ is a unitary, $\sum_{i=1}^n 
W_{ij}^* W_{ij}=I$, and we arrive at the relation
\begin{equation}\label{bazis}
\sum_{i,j=1}^n \Tr W_{ij} A W_{ij}^* B=\frac{n}{m}\Tr A \Tr B 
\end{equation}

We can transform this into another equivalent condition in terms of the left
multiplication and right multiplication operators. For $A,B \in   M_m(\bbbc)$, 
the operator $R_A$ is the right multiplication by $A$ and  $L_B$ is the left 
multiplication by $B$: $R_A, L_B: M_m(\bbbc)\to M_m(\bbbc),\, R_BX=XB,\, L_AX
=AX$. Equivalently, $L_A |e\>\<f|=|Ae\>\<f|$ and $R_B |e\>\<f|=|e\>\<B^*f|$.

The equivalent form of (\ref{bazis}) is the equation
$$
\frac{m}{n} \sum_{i,j=1}^n \<W_{ij}, R_A L_B  W_{ij}\>= \Tr A\, \Tr B= 
\Tr R_A L_B
$$
for every $A,B \in   M_m(\bbbc)$. Since the  operators
$R_A L_B$ linearly span the space of all linear operators on $M_m(\bbbc)$, 
$$
\frac{m}{n} \sum_{i,j=1}^n \Tr |W_{ij}\> \<W_{ij}| X =\frac{m}{n}
\sum_{i,j=1}^n \<W_{ij}, X  W_{ij}\>= \Tr X
$$
for every (super)operator $X:M_m(\bbbc)\to M_m(\bbbc)$. So we conclude
$$
\frac{m}{n}  \sum_{i,j=1}^n |W_{ij}\> \<W_{ij}|=I,
$$
where $I$ is the identity acting on the space $ M_m(\bbbc)$. \qed

Although the previous theorem is formulated for a tensor product, it covers the
general case. If $\iA_1$ is a subalgebra of $\iA$, $\iA_1 \iso M_n(\bbbc)$ and
$\iA \iso M_p(\bbbc)$, then $m:=p/n$ is an integer and $\iA \iso M_n(\bbbc)
\ot  M_m(\bbbc)$. (The subalgebra  $M_m(\bbbc)$ is the relative commutant of
$\iA_1$.)  Let us call the unitary satisfying the condition in the previous 
theorem as a {\bf useful unitary}. 

In the rest of the paper we work in the
situation $m=n$ and we denote the set of all $n^2\times n^2$ useful
unitaries by $\iM(n^2)$. To construct $k$ pairwise complementary subalgebras 
we need $k$ unitaries $W_1, W_2, \dots, W_k \in \iM(n^2)$ such that $W_1=I$ 
and $W_i W_j^*$ is a useful unitary if $i> j$.

Since $\< A, WBW^*\>=\< W^*AW, B\>$, we have $W \in \iM(n^2)$ if and only if 
$W^* \in \iM(n^2)$. This can be seen also from the condition of Theorem
\ref{T:2}.

\begin{Q} 
Let $\iA_1$ and  $\iA_2$ be subalgebras of $ M_n(\bbbc) \ot M_n(\bbbc)$
such that they isomorphic to $M_n(\bbbc)$. Set
$$
d:=\sup \{\tau(P_1 P_2): P_i {\rm \ is\ a\ minimal\ projection\ in\ }\iA_i\}.
$$
Then $d \le 1/n$. Assume that $1/n -d>0$.  Can we give a lower bound for
$$
S(\varphi|\iA_1)+S(\varphi|\iA_2) 
$$
as an analogue of the uncertainty relation in Theorem \ref{thm:16.2}?
\end{Q}

\begin{pl}
Now we generalize Example \ref{Pl:2}. We want to construct a unitary $W:= 
\sum_{ij} E_{ij} \ot W_{ij}$ such that $n^{-1/2}W_{ij}$ form an orthonormal basis
with respect to (\ref{E:HS}).

Let $X$ and $Y$ be the $n \times n$ unitaries from Example \ref{pl:Sch}, 
and let 
$(c_{ij})$ be a unitary such that $n |c_{ij}|^2=1$. Set
\begin{equation}\label{E:W}
W_{ij}:=c_{ij}X^{i}Z^{j}.
\end{equation}
Then
$$
\sum_j  W_{ij}(W_{kj})^*=\sum_j c_{ij}\overline{c}_{kj}X^{i-k}=\delta_{jk} I
$$
and $W$ is a unitary. Moreover, $\Tr W_{ij}^* W_{ij}= |c_{ij}|^2 \Tr I=1$.

In the case of $n=2$, $X=\sigma_1$ and $Z=\sigma_3$. Similarly to 
(\ref{E:Paulik}) we have the useful unitary
\begin{equation}\label{E:Paulik2}
W:=\frac{1}{\sqrt{2}}\left[\begin{array}{cc}
-\im\sigma_2 & \sigma_1 \\  \sigma_3 & I \end{array}\right].
\end{equation}
\qed \end{pl}
 
Since we have a unitary $W$ in $ M_n(\bbbc) \ot M_n(\bbbc)$ such that
it satisfies the condition of Theorem \ref{T:2}, we obtain examples of
complementary subalgebras.

\begin{pl}
Let $\iA$ be the algebra generated by the operators $a_1, a_1^*,a_2,a_2^*$
satisfying the {\bf canonical anticommutation relations}:
$$
\{a_1,a_1^*\}=\{a_2,a_2^*\}=I, 
\{a_1,a_1\}=\{a_1,a_2\}=\{a_1,a_2^*\}=\{a_2,a_2\}=0,
$$
where $\{A,B\}:=AB+BA$. Let $\iA_1$ be the subalgebra generated $a_1$ and 
$\iA_2$ be the subalgebra generated $a_2$. Then $\iA_1$ and $\iA_2$ are 
complementary. In the usual matrix representation 
$$
a_1=\left[\begin{array}{cc} 0 & 1 \\ 0 & 0 \end{array}\right]
\otimes \left[\begin{array}{cc} 1 & 0 \\ 0 & 1 \end{array}\right]
\quad \mbox{and}\quad
a_2=\left[\begin{array}{cc} 1 & 0 \\ 0 & -1 \end{array}\right]
\otimes \left[\begin{array}{cc} 0 & 1 \\ 0 & 0 \end{array}\right],
$$
therefore
$$
\iA_1=\left\{\left[\begin{array}{rrrr}
a & 0 & b & 0\\
0 & a & 0 & b\\
c & 0 & d & 0\\
0 & c & 0 & d\\
\end{array}\right]\right\},\quad
\iA_2=\left\{\left[\begin{array}{rrrr}
a & b & 0 & 0\\
c & d & 0 & 0\\
0 & 0 & a & -b\\
0 & 0 & -c & d\\
\end{array}\right]\right\}\,.
$$
The unitary sending $\iA_1$ to $\iA_2$ is
$$
V:=\left[\begin{array}{rrrr}
1 & 0 & 0 & 0\\
0 & 0 & 1 & 0\\
0 & -1 & 0 & 0\\
0 & 0 & 0 & 1\\
\end{array}\right]
$$
which is similar to $U$ in (\ref{E:LR}). The block matrix entries of $V$
form obviously a basis, so Theorem \ref{T:2} gives the complementarity.

More generally, consider the algebra $\iA$ generated by the operators 
$\{a_i: 1 \le i \le n\}$ satisfying the relations
\begin{eqnarray*}
a_i a_j +a_ja_i & = & 0 \\
a_i a_j^* +a_j^* a_i & = & \delta (i,j)
\end{eqnarray*}
for $1 \le i,j \le n$.  It is well-known that $\iA$ is isomorphic to the
algebra of $2^n \times 2^n$ matrices. Let $\{J_1,J_2\}$ be a partition of
the set $\{1,2,\dots ,n\}$ and let $\iA_j \subset \iA$ be the subalgebra
generated by $\{a_i: i \in J_j\}$, $j=1,2$. Since
\begin{equation}
\tau(ab)=\tau(a)\tau(b)
\end{equation}
holds for every $a \in \iA_1$ and $b \in \iA_2$, the subalgebras
$\iA_1$ and $\iA_2$ are complementary. (See \cite{AM, BR}.)\qed
\end{pl}

Assume that $\iA_1,\iA_2,\dots, \iA_m$ are complementary subalgebras 
of $M_n(\bbbc)\ot M_n(\bbbc)$ and each of them is isomorphic to  $M_n(\bbbc)$.
Since the dimension of $\iA_a \ominus \bbbc I$ is $n^2-1$, the inequality 
$n^4-1 \ge m(n^2-1)$ holds. This implies that $m \le n^2+1$. This trivial upper
bound is 5 for $n=2$. However, the maximum number of complementary subalgebras
is 4. This will be discussed in the next section.

\section{Two qubits}

We try to find a unitary $W$ again such that the subalgebra
$$
W\left[\begin{array}{cc} A & 0 \\ 0 & A \end{array}\right]W^*
\qquad (A \in  M_2(\bbbc))
$$
is complementary to $\bbbc I \ot M_2(\bbbc)$. The approach of the paper
\cite{nofive} is followed here. We may assume that $W$ has the 
{\bf Cartan decomposition} 
$$
W=(L_1 \ot L_2) N (L_3 \ot L_4)\,,
$$
where $L_1,L_2,L_3$ and $L_4$ are $2 \times 2 $ unitaries and
\begin{equation}\label{E:NA}
N= \exp( \alpha \im \,\sigma_1 \ot \sigma_1) 
\exp( \beta \im \,\sigma_2 \ot \sigma_2)
\exp( \gamma \im \,\sigma_3 \ot \sigma_3)
\end{equation}
is a $4 \times 4$ unitary in a special form, see equation (11) in \cite{Zhang}
or \cite{Da}. The subalgebra
$$
W( \bbbc I \ot M_2(\bbbc))W^*
$$
does not depend on $L_3$ and $L_4$, therefore we may assume that $L_3=L_4=I$.

The orthogonality of $\bbbc I \ot M_2(\bbbc)$ and $W( \bbbc I \ot 
M_2(\bbbc))W^*$ does not depend on $L_1$ and $L_2$. Therefore, the equations
$$
\Tr N(I \ot \sigma_i)N^*(I \ot \sigma_j)=0
$$
should be satisfied, $1 \le i,j \le 3$. We know from Theorem \ref{T:2} such 
that these conditions are equivalent to the property that the matrix elements 
of $N$ form a basis.

A simple computation gives that
$$
N=\sum_{i=0}^3 c_i \, \sigma_i \ot \sigma_i=
\left[\begin{array}{cccc}
c_0+c_3 & 0 & 0 & c_1-c_2\\
0 & c_0-c_3 & c_1+c_2 & 0\\
0 & c_1+c_2 & c_0-c_3 & 0\\
c_1-c_2 & 0 & 0 & c_0+c_3\\
\end{array}\right]\,,
$$
where
\begin{eqnarray*}
c_0&=&\cos \alpha\, \cos \beta \cos \gamma+ \im  \sin \alpha\, \sin \beta\, \sin \gamma \,,
\cr 
c_1 &=& \cos \alpha\, \sin \beta \sin \gamma + \im  \sin \alpha\, \cos \beta\, \cos \gamma\,,
\cr
c_2&=&  \sin \alpha\, \cos \beta \sin \gamma+ \im  \cos \alpha\, \sin \beta\, \cos \gamma\,,
\cr 
c_3&=& \sin \alpha\, \sin \beta \cos \gamma+ \im  \cos\alpha\, \cos \beta\, \sin \gamma\,.
\end{eqnarray*}
From the condition that the $2 \times 2$ blocks form a basis (see Theorem 
\ref{T:2}), we deduce the equations
$$
|c_0|^2=|c_1|^2=|c_2|^2=|c_3|^2=\frac{1}{4}
$$
and arrive at the following solution. Two of the values of
$\cos^2 \alpha, \cos^2 \beta$ and $ \cos^2 \gamma$ equal $1/2$
and the third one may be arbitrary. Let $\iN$ be the set of all
matrices such that the parameters $\alpha, \beta$ and $\gamma$ satisfy
the above condition, in other words two of the three values 
are of the form $\pi/4+k \pi/2$. ($k$ is an integer.) Let
\begin{equation}\label{E:N1}
\iN_1:= \{ N\in \iN: \alpha \mbox{\ is\ arbitrary,\ } \beta=\pi/4+k_1 \pi/2,
\mbox{\ and\ } \gamma=\pi/4+k_2 \pi/2\}
\end{equation}
and define $\iN_2$ and $\iN_3$ similarly. ($\iN=\iN_1 \cup \iN_2 
\cup \iN_3$.)

The conclusion of the above argument can be formulated as follows.

\begin{Thm}
$W \in \iM(4)$ if and only if $W=(L_1 \ot L_2)N(L_3 \ot L_4)$, where 
$L_i$ are $2 \times 2$ unitaries ($1\le i \le 4$) and $N \in \iN$.
\end{Thm}

It follows that $W \in \iM(4)$ if and only if $(U_1 \ot U_2)W \in \iM(4)$
for some (or all) unitaries $U_1$ and $U_2$. Note that this fact can be
deduced also from Theorem \ref{T:2}.

\begin{pl}\label{P:N_1}
A simple example for a unitary $N_3$ from $\iN_3$ is
$$
N_3= \left[\begin{array}{cccc}
1 & 0 & 0 & 0\\
0 & 0 & \im  & 0\\
0 & \im & 0 & 0\\
0 & 0 & 0 & 1\\
\end{array}\right]
$$
which corresponds to $\alpha=\beta=\pi/4$ and $\gamma=0$. One can check
that
\begin{eqnarray*}
N_3(I\ot \sigma_1)N_3^*&=&\sigma_2 \ot \sigma_3, \cr
N_3(I\ot \sigma_2)N_3^*&=&-\sigma_1 \ot \sigma_3, \cr
N_3(I\ot \sigma_3)N_3^*&=&\sigma_3 \ot I\,.
\end{eqnarray*} \qed
\end{pl}

A part of the example is true more generally \cite{Priko}:

\begin{lemma}
If $N_i \in \iN_i$, then $N_i(I\ot \sigma_i)N_i^*$ equals $\sigma_i \ot I$
up to a sign for $1 \le i \le 3$.
\end{lemma}

It is useful to know that for $N_1 \in \iN_1$, see (\ref{E:N1}), the
subalgebra $N_1 (\bbbc I \ot M_2(\bbbc))N_1^*$ does not depend on
the integers $k_1$ and $k_2$. Therefore, it is often convenient to
assume that $k_1=k_2=0$. Similar remarks hold for $\iN_2$ and $\iN_3$
\cite{Priko}.  

\begin{Thm}
Let $\iA^0\equiv \bbbc I \ot M_2(\bbbc)$ and $\iB\equiv M_2(\bbbc)\ot \bbbc I$.
Assume that the subalgebra $\iA^1 \subset M_2(\bbbc)\ot M_2(\bbbc)$ is isomorphic to
$M_2(\bbbc)$ and complementary to $\iA^0$. Then the intersection of $\iA^1$
and $\iB$ is not trivial.
\end{Thm}

\proof
There is a unitary $W=(L_1 \ot L_2)N$ such that $\iA^1=W \iA^0 W^*$, $L_1, L_2$
are $2 \times 2$ unitaries and $N \in \iM(4)$. Assume that $N \in \iN_i$. Then
$$
(L_1 \ot L_2)N(I \ot \sigma_i)N^*(L_1^* \ot L_2^*)=\pm L_1 \sigma_i L_1^* \ot I
$$
is in the intersection of $\iA^1$ and $\iB$ and $L_1 \sigma_i L_1^*$ cannot
be a constant multiple of $I$ (since its spectrum is $\{1,-1\}$). \qed 

This theorem implies that $M_2(\bbbc)\ot M_2(\bbbc)$ cannot contain 5 subalgebras
which are pairwise complementary and isomorphic to $M_2(\bbbc)$. The question
about the existence of 5 such subalgebras was raised in \cite{PHSz} and the answer
was given first in \cite{nofive}. The proof presented here is slightly different.

\begin{pl}
The $4 \times 4$ matrices 
$$
C= \left[\begin{array}{cccc}
a & 0 & 0 & b\\
0 & c & d & 0\\
0 & d & c & 0\\
b & 0 & 0 & a\\
\end{array}\right]
$$
form a commutative algebra $\iC$ isomorphic to $\bbbc^4$. Concretely,
the isomorphism $\kappa$ maps the above matrix into
$$
\kappa(c)=(a+b,a-b,c+d,c-d)\,.
$$
The spectral decomposition of $C$ is
$$
C=(a+b)P_+ +(a-b)P_- + (c+d)Q_++ (c-d)Q_-\,,
$$
where
$$
P_\pm =\frac{1}{2}\left[\begin{array}{cccc}
1 & 0 & 0 & \pm 1\\
0 & 0 & 0 & 0\\
0 & 0 & 0 & 0\\
\pm 1  & 0 & 0 & 1\\
\end{array}\right],\quad
Q_\pm =\frac{1}{2}\left[\begin{array}{cccc}
0 & 0 & 0 & 0\\
0 & 1 & \pm 1 & 0\\
0 & \pm 1 & 1 & 0\\
0  & 0 & 0 & 0\\
\end{array}\right]\,.
$$
Note that the projections $P_\pm$ and $Q_\pm$ correspond to the {\bf Bell 
basis}.

Let $E_\iC$ be the $\tau$-preserving conditional expectation onto the 
subalgebra $\iC$. This has the form
\begin{equation}\label{E:E}
E_\iC\Big(\sum_{ij} c_{ij}\, \sigma_i \ot \sigma_j\Big)=
\sum_{i} c_{ii}\,\sigma_i \ot \sigma_i\,.
\end{equation} 
It follows that
\begin{equation}\label{E:trC}
E_\iC (I \ot A)=E_\iC (A \ot I)= \tau(A)\,I.
\end{equation} \qed
\end{pl}
 
\begin{Thm} 
If $M\in \iC$ is a unitary, then the subalgebras $M(\bbbc I \ot 
M_2(\bbbc))M^*$
and $\iC$ are complementary. In particular, $\iC$ is complementary to
$\bbbc I \ot M_2(\bbbc)$ and $M_2(\bbbc) \ot \bbbc I$.
\end{Thm}

\proof
Assume that $A \in M_2(\bbbc)$ is traceless and $C \in \iC$. We have to
show that
$$
\Tr C^* M(I\ot A)M^* =0.
$$
This follows from (\ref{E:trC}):
\begin{eqnarray*}
\Tr C^* M(I\ot A)M^* &=&\Tr M^*C^* M(I\ot A)=\Tr E_\iC(M^*C^* M(I\ot A))
\cr &=&
\Tr M^*C^* M E_\iC(I\ot A)=\tau(A)\Tr M^*C^* M=0.
\end{eqnarray*}
\qed

The theorem tells us that a measurement corresponding to the Bell basis
is complementary to any local measurement of the two-qubit-system.

The algebra $M_2(\bbbc) \ot M_2(\bbbc)$ can be decomposed to complementary
subalgebras. Together with the identity, each of the following triplets 
linearly spans a subalgebra $\iA_j$ isomorphic to $M_2(\bbbc)$
($1 \le j \le 4$).
\begin{eqnarray*} 
&& \{ \sigma_0\otimes\sigma_1,\,\sigma_1\otimes\sigma_3,\,
\sigma_1\otimes\sigma_2\} \cr && 
\{\sigma_3\otimes\sigma_1,\,\sigma_1\otimes\sigma_1,\,
\sigma_2\otimes\sigma_0\} \cr && 
\{\sigma_1\otimes\sigma_0,\,\sigma_2\otimes\sigma_2,\,
\sigma_3\otimes\sigma_2\} \cr && 
\{\sigma_0\otimes\sigma_2,\,\sigma_2\otimes\sigma_3,\,
\sigma_2\otimes\sigma_1\}\,. 
\end{eqnarray*}
The orthogonal complement spanned by $\{\sigma_0 \ot \sigma_3,\sigma_3 \ot 
\sigma_0,\sigma_3 \ot \sigma_3\}$ is a commutative subalgebra.

\bigskip
{\bf Acknowledgement:} The author thanks to Paul Busch, Domenico 
D'Alessandro, Patrick Hayden, Pekka Lahti, Mil\'an Mosonyi and Andr\'as 
Sz\'ant\'o for discussions on related subjects.

\end{document}